\documentclass{ametsocV6.1}
\usepackage{enumitem}

\title{Understanding Stokes drift mechanism  via crest and trough phase estimates}

\authors{Anirban Guha\aff{a}\correspondingauthor{Anirban Guha, anirbanguha.ubc@gmail.com} 
 and Akanksha Gupta\aff{b} 
}
\affiliation{\aff{a}{School of Science and Engineering, University of Dundee, Dundee DD1 4HN, UK.}\\
\aff{b}{Scripps Institution of Oceanography, University of California, San Diego, CA, 92093, USA.}
}

\abstract{By providing mathematical estimates, this paper answers a fundamental question -- ``what leads to Stokes drift''? Although overwhelmingly understood for water waves, Stokes drift is a generic mechanism that stems from kinematics and occurs in any non-transverse wave in fluids. To showcase its generality, we undertake a comparative study of the pathline equation of sound (1D) and intermediate-depth water (2D) waves. Although we obtain a closed-form solution $\mathbf{x}(t)$ for the specific case of linear sound waves, a more generic and meaningful approach 
involves the application of asymptotic methods and expressing variables in terms of the Lagrangian phase $\theta$. We show that the latter reduces the  
2D pathline equation of water waves to 1D.
 Using asymptotic methods, we solve the respective pathline equation for sound and water waves, and for each case, we obtain a parametric representation of particle position $\mathbf{x}(\theta)$ and elapsed time $t(\theta)$. Such a  parametric description has allowed us to obtain second-order-accurate expressions for the time duration, horizontal displacement, and average horizontal velocity of a particle in the crest and trough phases. All these quantities are of higher magnitude in the crest phase in comparison to the trough, leading to a forward drift, i.e.\,Stokes drift.  We also explore particle trajectory due to second-order Stokes waves and compare it with linear waves. While finite amplitude waves modify the estimates obtained from linear waves, the understanding acquired from linear waves is generally found to be valid. }

\begin{document}

\maketitle

%
%
%
%
%
%

%

\section{Introduction}
\label{sec:intro}

In a generic way, Stokes drift is defined as the difference  between  wave-averaged velocity following a particle  (i.e.\,Lagrangian)   and that in a fixed reference frame (i.e.\,Eulerian) \citep{stokes1847,buhler2014waves,van2017stokes}:
\begin{align}
 \overline{\bf{u}^{\textnormal{SD}}}=\overline{\bf{u}^{\textnormal{L}}}-\overline{\bf{u}^{\textnormal{E}}}.
 \label{eq:SD_general}
\end{align} 
 Stokes drift mechanism has been particularly understood in the context of surface gravity waves (or simply, `water waves'), and is known to play a central role in the transport of various oceanic tracers, e.g.\,sediments, pollutants,  heat, nutrients, etc. { Detailed experimental results on Stokes drift show excellent match with the theory \citep{umeyama2012eulerian}.}
 The current understanding of the Stokes drift mechanism is succinctly given in \cite{van2017stokes}: ``\emph{A  fluid particle, which oscillates backwards and forwards due to the linear wave motion, spends more time in the forward-moving region underneath the crest than in the backwards-moving region underneath the trough and undergoes its forward motion at a greater height, where the velocities are larger.}'' A similar line of argument can also be found in other references \citep{craik1988wave}.
 In this regard, we raise a few fundamental questions:
 \begin{enumerate}[label=(\roman*)]
     \item Is it known  \emph{how long} does a particle spend underneath the crest in comparison to that underneath the trough?
     \item Stokes drift does exist in the truly 1D longitudinal wave problem \citep[\S10.1.1]{buhler2014waves}. Hence, the proposed argument behind Stokes drift mechanism -- forward motion happens because, at greater heights, velocities are larger -- although seems quite intuitive for water waves, does not hold for longitudinal (e.g.\,sound) waves, since there are no variations in the transverse direction. This naturally begs the question -- is there a generic explanation (including quantitative estimates) of the Stokes drift mechanism that is applicable to all kinds of non-transverse waves in fluids?

 \end{enumerate}
 To the best of our knowledge, answers to the above questions are not yet known. Hence even after 175 years of its discovery, some key aspects of the  Stokes drift mechanism have remained elusive. This paper aims to revisit the Stokes drift problem and develop a  mathematical framework to address such questions.

Classically, for evaluating Stokes drift, the wave-averaged Lagrangian velocity $\overline{\bf{u}^{\textnormal{L}}}$ in \eqref{eq:SD_general} is \emph{not} evaluated using a   Lagrangian (particle following) formalism, rather by undertaking a hybrid Eulerian-Lagrangian approach involving Taylor series expansion \citep{stokes1847}. 
For  small amplitude waves, this yields the following expression for Stokes drift at the second order of wave steepness:
\begin{equation}
\overline{\bf{u}^{\textnormal{SD}}}=\overline{\boldsymbol{\xi}^{(1)}\cdot \nabla \mathbf{u}^{(1)}},
\label{SD_classical}
\end{equation}
where $\mathbf{u}^{(1)}$ is the linear wave field (leading order in wave steepness), $\boldsymbol{\xi}^{(1)}$ is the corresponding linear displacement vector satisfying $\partial \boldsymbol{\xi}^{(1)} /\partial t\!=\!\mathbf{u}^{(1)}$ \citep{craik1988wave,van2017stokes}. 
This hybrid Eulerian-Lagrangian approach has been generalized and extended to finite amplitude waves via constructing the Generalized Lagrangian mean (GLM)  \citep{andrews1978exact}. GLM is a practical alternative to the  Lagrangian approach, which is often impractical in fluid dynamics problems \citep{buhler2014waves}. However, to answer the questions we seek, a  Lagrangian formalism is essential. 

By a  Lagrangian approach (referred to as the `exact theory' in   \cite{constantin2008particle}, hereafter CV08), we mean solving the pathline equation
of the form (2D case shown)
\begin{equation}
    \frac{d\boldsymbol{x}}{dt}=\boldsymbol{u}(\boldsymbol{x},t)=\underbrace{\boldsymbol{u}^{(1)}(kx-\omega t,z)}_{\textnormal{Linear term}} + \underbrace{\boldsymbol{u}^{(2)}(2kx-2\omega t,z)+\ldots}_{\textnormal{Higher-order wave componenets}}
    \label{eq:the_key}
\end{equation}
\emph{without} invoking the `small excursion' approximation (which uses Taylor series expansion). { The R.H.S. denotes a regular wave of permanent form};  $\omega$ and $k$  are respectively the  frequency and wavenumber, and
$\boldsymbol{u}^{(1)}, \boldsymbol{u}^{(2)},\ldots$ represent {terms in the perturbation series  of $\boldsymbol{u}$ (or more precisely,  `Stokes expansion') such that} 
 $|\boldsymbol{u}^{(1)}|\gg |\boldsymbol{u}^{(2)}|\gg\ldots$ CV08 attempted to exactly solve \eqref{eq:the_key} for linear water waves ({keeping in mind that \eqref{eq:the_key} itself is an asymptotic expansion}). The above pathline equation for linear water waves represents a 2D nonlinear dynamical system, finding the exact solution for which is probably impossible (which is the reason why  \cite{stokes1847} employed the hybrid Eulerian-Lagrangian approach in the first place). Hence, the mathematical proofs provided in CV08, with related findings reported in their earlier work \citep{constantin2006trajectories},   are in the form of inequalities, and not equations/expressions. Using phase plane analysis, CV08 found three key results:  (a) a particle takes more time to complete its $2\pi$ phase cycle than the wave period (which corroborates  \cite{longuet1986eulerian}), (b) particle trajectories under linear water waves are open, 
 a consequence of which is that (c) particles undergo a slow drift in the direction of wave propagation. 
In this paper, we build on the understanding provided in CV08 {by undertaking a different approach which combines Lagrangian phase parameterization and asymptotic expansions.}  Through this, we   solve \eqref{eq:the_key} for longitudinal (e.g.\,sound) waves and intermediate depth water waves,
 which  eventually provide answers to questions like (i)--(ii).  

\section{The  1D problem: longitudinal waves}
\label{sec:sound}
The velocity field induced by  the passage of small amplitude longitudinal waves  in a fluid can be written in the following generic way:
\begin{align}
u(x,t)= \epsilon c  \cos (kx-\omega t),
 \label{eq:sound1}
\end{align}
where $\epsilon\!\ll\!1$, $c\!=\!\omega/k$ is the wave speed.  For sound waves, $\epsilon=\mathcal{P}/\rho_0 c^2$, where $\mathcal{P}$ is the amplitude of (isentropic) pressure fluctuation, and $\rho_0$ is the reference density.
 In fact, an isentropic system demands that an acoustic wave of permanent form must be linear \citep{earnshaw1860,falkovich2011fluid}. 

Equation \eqref{eq:sound1} represents a velocity field, hence by definition is an Eulerian formalism. In order
to map particle trajectories in this flow field, the same problem needs to be interpreted from the Lagrangian perspective, leading to the 1D pathline equation:
\begin{align}
 \frac{\mathrm{d} x}{\mathrm{d} t}  =u(x(t),t)= \epsilon c \cos (kx(t)-\omega t).
 \label{eq:sound11}
\end{align}
The above equation, arguably the simplest representation of the motion of a particle in a wave field, does reveal the important fact that the dynamical system is  \emph{nonlinear} (and non-autonomous), although it is derived from   \emph{linear} wave theory. The closed-form solution of \eqref{eq:sound11} is
{
\begin{align}
x(t)=c t -\frac{2}{k}\tan^{-1}\Biggl[\sqrt\frac{1-\epsilon}{1+\epsilon}\tan\bigg\{\frac{\omega\sqrt{1-\epsilon^2}}{2}t-\Phi_0\bigg\}\Biggr]\,\,\,\,\textnormal{where}\,\,\,\,
\Phi_0=\tan^{-1}\bigg[\sqrt\frac{1+\epsilon}{1-\epsilon}\tan\bigg(\frac{kx_0}{2}\bigg)\bigg]
\label{eq:exact_sound}
\end{align}
in which  $x_0\equiv x(0)$. To the best of our knowledge, the above expression has not been previously reported in the literature. Equation \eqref{eq:exact_sound} straightforwardly shows that the integral curve $x(t)$ is \emph{not} time-periodic, but does contain periodic (tangent) function. Another  key insight is naturally revealed from \eqref{eq:exact_sound} --
    the relevant time period is
    \begin{equation}
        T^{\textnormal{L}}=\frac{T^{\textnormal{E}}}{\sqrt{1-\epsilon^2}}\approx \bigg(1+
    \frac{\epsilon^2}{2}\bigg)T^{\textnormal{E}},
    \label{eq:exact_sound_time}
    \end{equation}
  where   $T^{\textnormal{E}}=2\pi/\omega$ is the wave period i.e.\,time period in the classical Eulerian sense. $T^{\textnormal{L}}$ denotes the Lagrangian time period, the meaning of which will be revealed shortly.}
{Furthermore, \eqref{eq:exact_sound} also reveals Stokes drift. Using $\sqrt{(1-\epsilon)/(1+\epsilon)}\approx 1$, we find that a particle undergoes  a net forward drift after a period of $T^{\textnormal{L}}$:
\begin{equation}
    x(T^{\textnormal{L}})-x_0\approx cT^{\textnormal{L}}-\frac{2}{k} \bigg(\pi-\frac{kx_0}{2} \bigg) -x_0\approx\frac{\pi}{k}\epsilon^2.
    \label{eq:exact_sound_drift}
\end{equation}
}
{While this particular case of linear longitudinal waves yields a closed-form solution, the same cannot be expected for waves whose pathline equation represents a 2D or a 3D dynamical system (e.g.\,water waves). This necessitates a more generic mathematical technique, e.g.\,the hybrid Eulerian-Lagrangian approach. }
Here,  $x(t)$ in the R.H.S. of \eqref{eq:sound11} is replaced by a fixed value  $\tilde{x}_0$. In other words, the phase part is treated as Eulerian. Using Taylor-series expansion discussed in \S\ref{sec:intro}, we obtain \citep{falkovich2011fluid}:
\begin{equation}
    x(t)=\tilde{x}_0-\frac{\epsilon}{k}\sin(k\tilde{x}_0-\omega t) - \frac{\epsilon^2}{4k}\sin2(k\tilde{x}_0-\omega t)+\frac{\epsilon^2 c}{2}t.
    \label{eq:hyb_lag_eul_sound}
\end{equation}
{The graph of the above expression shows an excellent match with the exact solution \eqref{eq:exact_sound}. However, the two expressions have a key difference which will become a central aspect in our study -- the relevant time period in \eqref{eq:hyb_lag_eul_sound} is $T^{\textnormal{E}}$ and not $T^{\textnormal{L}}$.}  
The net forward drift that a particle undergoes after one wave period is found to be $\epsilon^2 \pi/k$, matching \eqref{eq:exact_sound_drift}.
However from  \eqref{eq:exact_sound_drift} or \eqref{eq:hyb_lag_eul_sound}, it is not clear what led to this drift mechanism.   

In order to obtain a deeper insight into the drift mechanism, we express \eqref{eq:sound1} in  terms of the `Lagrangian' phase  
$\theta(t)\equiv\omega t -kx(t)$: 
\begin{equation}
 \frac{\mathrm{d} \theta}{\mathrm{d} t}=\omega(1-\epsilon \cos \theta).
 \label{eq:sound2}
\end{equation}
This change in variable 
renders the dynamical system autonomous.   
The quantity `$\epsilon \cos \theta$' in the R.H.S. of \eqref{eq:sound2}  introduces aperiodicity in the system. Instead of obtaining a closed-form solution like \eqref{eq:exact_sound}, we apply asymptotic power series expansion in $\epsilon$, which has the advantage of being extendable to higher dimensional systems:
\begin{align} 
\int \, dt &=\frac{1}{\omega}\int_{} \frac{d\theta}{1-\epsilon \cos \theta}  \approx \frac{1}{\omega}\int \,(1+\epsilon \cos \theta + \epsilon^2 \cos^2\theta  +\ldots)\,d\theta  \nonumber \\
 \Rightarrow  t & = 
\frac{\theta}{\omega}+\frac{\sin \theta}{\omega} \epsilon +\frac{2\theta+ \sin 2\theta}{4\omega} \epsilon^2, 
 \label{eq:sound3}
\end{align}
 It is obvious that $\theta$ (and not $t$) is the independent variable, which would allow us to calculate, for example, the time elapsed between two events (e.g. when $\theta=0$ and $\theta=2\pi$). 
The crest and trough phases of a cosine waveform are as follows:
\begin{equation}
  \textnormal{Crest phase}: \theta \in \bigg[0,\frac{\pi}{2}\bigg]\cup \bigg[\frac{3\pi}{2},2\pi\bigg],\qquad    \textnormal{Trough phase}: \theta \in \bigg[\frac{\pi}{2},\frac{3\pi}{2}\bigg].
\label{eq:linear_phase_relation}
\end{equation}
Applying this in \eqref{eq:sound3}, we readily obtain
\begin{subequations}
\begin{align}
   &\textnormal{Crest phase duration}:\,\,\,\,\,\,\,\,\,T^{\textnormal{L},\textnormal{crest}}= \bigg(\frac{1}{2}+
   \frac{\epsilon}{\pi}+\frac{\epsilon^2}{4}\bigg)T^{\textnormal{E}}, \label{eq:T_crest_sound}\\
    &\textnormal{Trough phase duration}:T^{\textnormal{L},\textnormal{trough}}=\bigg(\frac{1}{2}-
   \frac{\epsilon}{\pi}+\frac{\epsilon^2}{4}\bigg)T^{\textnormal{E}}. \label{eq:T_trough_sound}
\end{align}
\end{subequations}
  Hence during its phase cycle of $2\pi$,  a particle spends an amount of time that is greater (lesser) than half of the wave period in the crest (trough) segment. In fact, a particle spends $2 T^{\textnormal{E}} \epsilon/\pi$  more time in the crest segment than in the trough segment. 

Adding \eqref{eq:T_crest_sound} and \eqref{eq:T_trough_sound} yields the Lagrangian time period
\begin{equation}
    T^{\textnormal{L}}= T^{\textnormal{L},\textnormal{crest}}+T^{\textnormal{L},\textnormal{trough}}=\bigg(1+
    \frac{\epsilon^2}{2}\bigg)T^{\textnormal{E}}>T^{\textnormal{E}}, 
    \label{eq:sound4}
\end{equation}
which matches the expression in \eqref{eq:exact_sound_time}.  $T^{\textnormal{L}}$, which naturally appears in the closed-form solution \eqref{eq:exact_sound}, is therefore a  measure of the time taken by a particle to complete its $2\pi$ phase cycle.  Hence a particle takes an additional $T^{\textnormal{E}}\epsilon^2/2$  time to complete its $2\pi$ phase cycle than its `carrier' longitudinal wave to complete one wave period.

The drift of a particle follows directly from \eqref{eq:sound3} after using the fact that $kx=\omega t -\theta$:
\begin{equation}
     x =\frac{\sin \theta}{k} \epsilon +\frac{2\theta+ \sin 2\theta}{4k} \epsilon^2.
    \label{eq:sound51}
\end{equation}
Particle displacement during the crest and trough segments of its phase cycle  is straightforwardly obtained from \eqref{eq:sound51}:
\begin{subequations}
\begin{align}
&\textnormal{Crest phase displacement}:\,\,\,\,\,\,\,\,x^{\textnormal{L,crest}}  = \bigg(\frac{\epsilon}{\pi}+\frac{\epsilon^2}{4} \bigg)\frac{2\pi}{k},  \label{eq:x_crest_sound}\\
& \textnormal{Trough phase displacement}:x^{\textnormal{L,trough}} = \bigg(\!-
   \frac{\epsilon}{\pi}+\frac{\epsilon^2}{4} \bigg)\frac{2\pi}{k}. \label{eq:x_trough_sound}
\end{align}
\end{subequations}
 Equations \eqref{eq:x_crest_sound}--\eqref{eq:x_trough_sound} reveal that a particle undergoes greater displacement in the crest phase than that in the trough phase. Therefore after one phase cycle, there is a net  displacement $x^{\textnormal{L,net}}$ of a particle in the direction of wave propagation: 
\begin{align}   x^{\textnormal{L,net}}=x^{\textnormal{L,crest}}+x^{\textnormal{L,trough}}= \frac{\pi}{k}\epsilon^2.
\label{eq:xsd_sound}
\end{align}
This exactly matches the expression obtained in \eqref{eq:exact_sound_drift} or \eqref{eq:hyb_lag_eul_sound}. 
We can also  evaluate the {average particle velocities in the crest and trough phases}:
\begin{subequations}
\begin{align}
& \textrm{{Avg. particle velocity in crest phase}:\,\,\,\,\,\,\,\,\,\,\,} \overline{u^{\textnormal{L,crest}}}=\frac{x^{\textnormal{L,crest}}}{T^{\textnormal{L,crest}}}  = \frac{2}{\pi}c\epsilon+\frac{c}{2}\bigg[1-\frac{8}{\pi^2}\bigg]\epsilon^2,\\
& \textrm{{Avg. particle velocity in trough phase}:\,\,\,} \overline{u^{\textnormal{L,trough}}} =\frac{x^{\textnormal{L,trough}}}{T^{\textnormal{L,trough}}} = -\frac{2}{\pi}c\epsilon+\frac{c}{2}\bigg[1-\frac{8}{\pi^2}\bigg]\epsilon^2.
\end{align}
\end{subequations}
 Hence the average particle velocity is faster 
in the crest phase  than that in the trough:
\begin{equation}
   |\,\overline{u^{\textnormal{L,crest}}}\,|-|\,\overline{u^{\textnormal{L,trough}}}\,|=c\bigg(1-\frac{8}{\pi^2}\bigg)\epsilon^2>0.
    \label{eq:avg_vel_sound}
\end{equation}


Finally, the  drift velocity is given by
\begin{equation} \overline{u^{\textnormal{SD}}}=\frac{x^{\textnormal{L,net}}}{T^{\textnormal{L}}}=\frac{c}{2}\epsilon^2.
    \label{eq:sound_SD}
\end{equation} 
The above expression signifies Stokes drift for all kinds of longitudinal waves in fluids, including sound waves and shallow water waves. To verify this, we also evaluate Stokes drift via the traditional route, i.e.\,\eqref{SD_classical}, which yields the exact same expression c.f.\,\citet[(10.12)]{buhler2014waves}. Furthermore,  the drift velocity can also be computed following the formal definition in \eqref{eq:SD_general}: $\overline{u^{\textnormal{SD}}}=\overline{u^{\textnormal{L}}}-\overline{u^{\textnormal{E}}}$.  As previously mentioned, we always undertake a  Lagrangian formalism, hence the  Lagrangian mean velocity is evaluated as follows:
\begin{align}
 \overline{u^{\textnormal{L}}}=\frac{1}{T^{\textnormal{L}}}\int_{0}^{T^{\textnormal{L}}} u(x(t),t) \,dt= \frac{\epsilon}{T^{\textnormal{L}}k}\int_{0}^{2 \pi} \frac{\cos \theta}{1-\epsilon \cos\theta}  \, d\theta=\frac{c }{2}\epsilon^2.
 \label{eq:lag_mean_sound}
\end{align}
Notice that in \eqref{eq:lag_mean_sound}, averaging is performed over ${T^{\textnormal{L}}}$ and not ${T^{\textnormal{E}}}$, since the former is the correct choice for a  Lagrangian approach. The Eulerian mean   velocity is given by
\begin{align}
 \overline{u^{\textnormal{E}}}=\frac{1}{T^{\textnormal{E}}}\int_{0}^{T^{\textnormal{E}}} u(x,t) \,dt= \frac{c }{T^{\textnormal{E}}} \epsilon  \int_{0}^{T^{\textnormal{E}}}  \cos (kx-\omega t)\,dt=0.
 \label{eq:Eul_mean_sound}
\end{align}
Therefore $\overline{u^{\textnormal{SD}}}$, obtained by subtracting  \eqref{eq:Eul_mean_sound} from \eqref{eq:lag_mean_sound}, exactly matches \eqref{eq:sound_SD}. 

To summarize, by introducing Lagrangian phase parameterization,  we have provided second-order accurate asymptotic estimates to support the following statement in \citet[\S10.1.1]{buhler2014waves}: ``\ldots\emph{in this longitudinal wave a fluid particle that is moved back
and forth by the linear wave is spending slightly more time in the forward-moving phase region than in the backward-moving one}''. By evaluating $T^{\textnormal{L,crest}}$, $T^{\textnormal{L,trough}}$, $x^{\textnormal{L,crest}}$, $x^{\textnormal{L,trough}}$, $\overline{u^{\textnormal{L,crest}}}$, and $\overline{u^{\textnormal{L,trough}}}$, we have proved that Stokes drift in longitudinal waves arise because a particle in the wave field spends more time,  undergoes greater horizontal displacement, and travels at a faster average horizontal velocity in the crest phase in comparison to the trough phase. 
Note that the hybrid Eulerian-Lagrangian approach provides an accurate estimate of Stokes drift (obtained by averaging over one wave period), but does not yield much physical insight into the drift mechanism. Such an issue stems from the fact that the hybrid approach treats phase as Eulerian, and as a consequence, cannot distil the variations between the crest and the trough phases, which is essential to quantitatively understand the drift mechanism. 
{Furthermore, while the closed-form solution (which we were able to derive for this particular problem) highlights the significance of the Lagrangian time period and also reveals Stokes drift, it lacks generality and does not straightforwardly provide any additional insights.}

\section{The 2D problem: intermediate depth water waves}
\label{sec:water}

\subsection{Linear Waves:}

Under the realm of 2D potential flow theory, the surface elevation $\eta$ and the corresponding velocity potential $\phi$ due to progressive, linear water waves in a fluid of constant depth $H$ is 
given as follows: 
\begin{subequations}
\begin{align}
\eta(x,t) &= \frac{\epsilon}{k} \cos(k x  -\omega t),\label{eq:eta_wat}\\
\phi(x,z,t)&=\frac{\epsilon c}{k} \dfrac{\cosh k(z+H)}{\sinh \alpha}  \sin (k x-\omega t),\label{eq:phi_wat}
\end{align}
\end{subequations}
where $\alpha=kH$ is the nondimensional wavenumber;
$\epsilon\!=\!ka\!\ll\!1$, and $\epsilon/\alpha=a/H\ll 1$, where $a$ denotes wave amplitude.
The velocity field can be straightforwardly obtained from the velocity potential in \eqref{eq:phi_wat}:  $(u,w)=(\partial\phi/\partial x,\partial\phi/\partial z)$. The problem now needs to be interpreted from the Lagrangian perspective for obtaining the pathline equation: 
\begin{subequations}
\begin{align}
 \frac{\mathrm{d} x}{\mathrm{d} t}=u(x(t),z(t),t) &= \epsilon c \dfrac{\cosh k(z(t)+H)}{\sinh \alpha}  \cos (k x(t)-\omega t), \label{eq:pathline_x}\\
\frac{\mathrm{d} z}{\mathrm{d} t}= w(x(t),z(t),t)  & =  \epsilon c \dfrac{\sinh k(z(t)+H)}{\sinh \alpha}  \sin (k x(t)-\omega t). \label{eq:pathline_z}
\end{align}
\end{subequations} 
Equations \eqref{eq:pathline_x}--\eqref{eq:pathline_z} represent a 2D, nonlinear, and non-autonomous dynamical system stemming from linear wave theory, and can be regarded as a 2D analog of the 1D pathline equation for linear longitudinal waves given in \eqref{eq:sound11}. Unlike the longitudinal wave problem in \S\ref{sec:sound}, we cannot find a closed-form solution for the integral curve $(x(t),z(t))$. However, the mathematical strategy developed in \S\ref{sec:sound} is directly applicable to water waves as well. In this regard, the first step is to substitute  $\theta(t)=\omega t -kx(t)$ in \eqref{eq:pathline_x}--\eqref{eq:pathline_z} to make the system of equations autonomous. 
{In addition,  we also make use of the kinematic boundary condition, which essentially implies that a particle has to forever remain on the material surface it is initially located.} Therefore, a particle on the free surface must satisfy $z=\eta=(\epsilon/k)\cos{\theta}$. As a result, the R.H.S. of \eqref{eq:pathline_x}--\eqref{eq:pathline_z} can be   expressed \emph{solely} as a function of $\theta$:
\begin{subequations}
\begin{align}
\frac{d \theta}{d t} &=\omega\bigg[1-\epsilon \frac{\cosh (\epsilon \cos \theta+\alpha)}{\sinh \alpha}  \cos \theta \bigg], \label{eq:pathline_theta}\\
w(\theta) & = -\epsilon c \dfrac{\sinh (\epsilon \cos \theta+\alpha)}{\sinh \alpha}  \sin \theta. \label{eq:pathline_z2}
\end{align}
\end{subequations}
Expressing in terms of $\theta$ yields a major advantage  -- \emph{the 2D water-wave pathline equation has essentially become  1D}. In other words, particle trajectory due to linear water waves is fundamentally governed by \eqref{eq:pathline_theta} -- a first-order, nonlinear ODE in $\theta$,  which is mathematically analogous (albeit, more involved) to that of \eqref{eq:sound2} for longitudinal waves. Equation \eqref{eq:pathline_theta} will be the cornerstone of all subsequent derivations,  the only use of \eqref{eq:pathline_z2} will be in the evaluation of the average vertical velocity (which must be zero, as shown later in \eqref{eq:wL_water}--\eqref{eq:wE_water}). We note in passing that the $z$-pathline equation, \eqref{eq:pathline_z}, can also be used to derive an expression for $d\theta/dt$ that is asymptotic to \eqref{eq:pathline_theta}, see Appendix A.
Moreover, while the analysis that will follow from \eqref{eq:pathline_theta}--\eqref{eq:pathline_z2} will be based on a particle at the free surface, the entire procedure can be straightforwardly generalized to a particle on a material surface at a given depth inside the fluid domain, see Appendix B. 
 
 Here we also point out the key difference between \eqref{eq:pathline_theta}--\eqref{eq:pathline_z2} and the autonomous pathline equation, which is (3.6) in CV08. While we both start from \eqref{eq:pathline_x}--\eqref{eq:pathline_z},  CV08 didn't use a substitution for $z$ in  \eqref{eq:pathline_x}--\eqref{eq:pathline_z}, hence their resultant autonomous dynamical system always remain 2D (i.e. in $\theta$--$z$, when written in our variables). While CV08 made the crucial proof (in terms of inequalities) that \eqref{eq:pathline_x}--\eqref{eq:pathline_z} does not have periodic solutions, our approach will consolidate their findings, and in addition, will provide various mathematical estimates like $x^{\textnormal{L,crest}}$, $x^{\textnormal{L,trough}}$, $T^{\textnormal{L,crest}}$, $T^{\textnormal{L,trough}}$, $\overline{u^{\textnormal{L}}}$, $\overline{u^{\textnormal{SD}}}$, etc.


Making use of the following approximations: 
\[\cosh(\epsilon \cos \theta)=1+(1/2) \epsilon^2\cos^2 \theta+\mathrm{O}(\epsilon^4)\,\, \mathrm{and}\,\,  \sinh(\epsilon \cos \theta)= \epsilon \cos \theta+\mathrm{O}(\epsilon^3),\]
followed by asymptotic power series expansion, \eqref{eq:pathline_theta} on integration yields
\begin{align}
t&= \frac{\theta}{\omega}+\frac{\mathcal{C}_{\alpha} \sin\theta}{\omega}  \epsilon+\frac{(1 +\mathcal{C}_{\alpha}^2) \left(2\theta+\mathcal{B}\sin 2 \theta \right)}{4\omega}  \epsilon^2, \label{eq:phase_vs_time_water} 
\end{align}
where $\mathcal{C}_{\alpha} = \coth \alpha$. Moreover  $\mathcal{B}=1$, which is \emph{incorrect}. The error in estimating $\mathcal{B}$ arises  because we neglected the $\mathrm{O}(\epsilon^2)$ term in 
\eqref{eq:eta_wat}--\eqref{eq:phi_wat}, but expressed \eqref{eq:phase_vs_time_water} correct up to $\mathrm{O}(\epsilon^2)$. 
The correct value of $\mathcal{B}$ is given in \eqref{appeq:water_finite_2}, and is obtained via a consistent asymptotic treatment by employing the  second-order Stokes  wave theory; see \S\ref{sec:water}\ref{Appsubsec:water}.  We will shortly see that $\mathcal{B}$ has no significance in the quantities we seek.  

\begin{figure}
    \centering
   \includegraphics[width=0.8\textwidth]{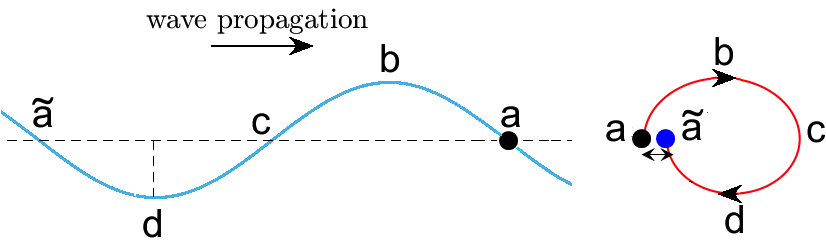}
    \caption{A rightward propagating linear water wave. A particle initially located at `a' traverses an open trajectory  a--b--c--d--$\Tilde{\mathrm{a}}$ in the clockwise direction. The endpoint $\Tilde{\mathrm{a}}$ signifies completion of $2\pi$ rotation. Each point marked on the particle trajectory corresponds to that on the wave. 
    The linear distance between `a' and `$\Tilde{\mathrm{a}}$' is  the Stokes correction $x^{\textnormal{L,net}}$.     }
    \label{fig:sch_waterwave}
\end{figure}

From \eqref{eq:phase_vs_time_water}, the time taken by a particle under  the crest (a--b--c) and trough (c--d--$\Tilde{\mathrm{a}}$) segments of a water wave,  shown in Fig.\,\ref{fig:sch_waterwave}, are respectively given by 
\vspace{-0.1cm}
\begin{subequations}
\begin{align} 
 & \textrm{Crest phase duration:\,\,\,\,\,\,\,\,\,\,} T^{\textnormal{L,crest}}= \bigg[\frac{1}{2}+
   \frac{\epsilon}{\pi}\mathcal{C}_{\alpha}+\frac{\epsilon^2}{4}(1+\mathcal{C}_{\alpha}^2)\bigg]
   T^{\textnormal{E}},\label{eq:T_crest_water}
   \\  
 & \textrm{Trough phase duration:\,\,}T^{\textnormal{L,trough}}=
  \bigg[\frac{1}{2}-
   \frac{\epsilon}{\pi}\mathcal{C}_{\alpha}+\frac{\epsilon^2}{4}(1+\mathcal{C}_{\alpha}^2)\bigg]
   T^{\textnormal{E}}.
 \label{eq:T_trough_water}
\end{align}
\end{subequations}
For deriving \eqref{eq:T_crest_water}--\eqref{eq:T_trough_water}, we have followed the same procedure that was used to derive \eqref{eq:T_crest_sound}--\eqref{eq:T_trough_sound}.
 Hence, similar to that of sound waves,  a particle while completing its phase cycle of $2\pi$, spends more (less) than half of the wave period in the crest (trough) segment. Precisely, a particle spends $2T^{\textnormal{E}}\mathcal{C}_{\alpha}\epsilon/\pi$ 
 more time in the crest segment than in the trough segment. We also note that \eqref{eq:T_crest_water}--\eqref{eq:T_trough_water} is independent of $\mathcal{B}$. The same will be observed while evaluating $x^{\textnormal{L,crest}}$, $x^{\textnormal{L,trough}}$, $\overline{u^{\textnormal{L,crest}}}$, and $\overline{u^{\textnormal{L,trough}}}$.


The total time taken by a particle to complete one phase cycle is 
\begin{align}
T^{\textnormal{L}}=T^{\textnormal{L,crest}}+T^{\textnormal{L,trough}}=\bigg[1+\frac{\epsilon^2}{2}(1+\mathcal{C}_{\alpha}^2)\bigg]T^{\textnormal{E}}. 
\label{eq:delta_T_general}
\end{align}
 Analogous to longitudinal waves,  \eqref{eq:delta_T_general} shows that the Lagrangian time period  $T^{\textnormal{L}}$ is more than the wave time period $T^{\textnormal{E}}$ by $  (1+\mathcal{C}_{\alpha}^2)T^{\textnormal{E}} \epsilon^2/2$. In practical terms, $T^{\textnormal{L}}$ would be the period measured by floating buoys in oceans, which are transported by water waves.  
 In fact, an expression similar to  \eqref{eq:delta_T_general} was provided in \citet[(2.3)]{longuet1986eulerian} usung first principles.
While the  the expression in  Longuet-Higgins  \emph{validates} our derivation, we emphasize  that following his procedure, it is \emph{not} possible to obtain quantities like $T^{\textnormal{L,crest}}$, $T^{\textnormal{L,trough}}$, $x^{\textnormal{L,crest}}$, $x^{\textnormal{L,trough}}$, $\overline{u^{\textnormal{L,crest}}}$, and $\overline{u^{\textnormal{L,trough}}}$.


We follow the same procedure as in \S\ref{sec:sound} to express $x$ in terms of $\theta$:
\begin{equation}
x=\frac{\mathcal{C}_{\alpha} \sin\theta}{k}  \epsilon+\frac{(1 +\mathcal{C}_{\alpha}^2) \left(2\theta+\mathcal{B}\sin 2 \theta \right)}{4k}  \epsilon^2. \label{eq:x_waterwaves}
\end{equation}
Particle displacement under the crest and trough segments can be calculated from \eqref{eq:x_waterwaves}:
\begin{subequations}
\begin{align}
& \textrm{Crest phase displacement:\,\,\,\,\,\,\,\,\,\,\,\,} x^{\textnormal{L,crest}}  = \bigg[\frac{\epsilon}{\pi}\mathcal{C}_{\alpha}+\frac{\epsilon^2}{4}(1+\mathcal{C}_{\alpha}^2)\bigg]\frac{2\pi}{k},\label{eq:water_x_crest}\\
& \textrm{Trough phase displacement:\,\,\,} x^{\textnormal{L,trough}}  = \bigg[\!-
   \frac{\epsilon}{\pi}\mathcal{C}_{\alpha}+\frac{\epsilon^2}{4}(1+\mathcal{C}_{\alpha}^2)\bigg]\frac{2\pi}{k}.\label{eq:water_x_trough}
\end{align}
\end{subequations}
Therefore, a particle undergoes greater displacement in the crest phase as compared to the trough phase, which in turn implies that the particle will drift in the direction of wave propagation. The net displacement, shown in Fig.\,\ref{fig:sch_waterwave}, is given as follows:
\begin{align}
     x^{\textnormal{L,net}}=x^{\textnormal{L,crest}}+x^{\textnormal{L,trough}}=x|^{a-c}- x|^{\Tilde{\mathrm{a}}-c}= \frac{\pi (1+\mathcal{C}_{\alpha}^2)}{k} \epsilon^2.
\end{align}

The  {average horizontal velocities in the crest and trough phase}  are given as follows:
\begin{subequations}
\begin{align}
& \textrm{{Avg. particle velocity in crest phase}:\,\,\,\,\,\,\,\,\,\,\,} \overline{u^{\textnormal{L,crest}}}  = \frac{2\mathcal{C}_{\alpha}}{\pi}c\epsilon+\frac{c}{2}\bigg[1+\mathcal{C}_{\alpha}^2\bigg(1-\frac{8}{\pi^2}\bigg)\bigg]\epsilon^2,\label{eq:water_u_av_crest}\\
& \textrm{{Avg. particle velocity in trough phase}:\,\,\,} \overline{u^{\textnormal{L,trough}}}  =-\frac{2\mathcal{C}_{\alpha}}{\pi}c\epsilon+\frac{c}{2}\bigg[1+\mathcal{C}_{\alpha}^2\bigg(1-\frac{8}{\pi^2}\bigg)\bigg]\epsilon^2. \label{eq:water_u_av_trough}
\end{align}
\end{subequations}
Hence the particle's average horizontal  velocity in the crest phase is faster than that in the trough phase: 
\begin{equation}
   |\,\overline{u^{\textnormal{L,crest}}}\,|-|\,\overline{u^{\textnormal{L,trough}}}\,|=c\bigg[1+\mathcal{C}_{\alpha}^2\bigg(1-\frac{8}{\pi^2}\bigg)\bigg]\epsilon^2>0.
    \label{eq:avg_vel_waterwave}
\end{equation}

\subsection{Second-order Stokes waves}
\label{Appsubsec:water}
As already discussed, the linear velocity field is order-wise inconsistent in evaluating $\mathrm{O}(\epsilon^2)$ estimates. Hence for legitimate results, a comparison should be made with the formally consistent second-order Stokes wave theory. 
The expression for surface elevation in the second-order Stokes wave theory reads \citep{dingemans1997water}:
\begin{equation}
    \eta(x,t) = \frac{\epsilon}{k} \cos(k x  -\omega t) + \frac{\epsilon^2}{k} \Gamma \cos2(k x  -\omega t), \label{appeq:eta}
\end{equation}
where $\Gamma\!=\!{(3-\tanh^2\alpha)}/{(4\tanh^3\alpha)}$.
The second-order component impacts the phases as follows: 
\begin{equation}
  \textnormal{Crest phase}: \theta\! \in \!\bigg[0,\frac{\pi}{2}-\epsilon\Gamma\bigg]\cup \bigg[\frac{3\pi}{2}+\epsilon\Gamma,2\pi\bigg],\,\,    \textnormal{Trough phase}: \theta \!\in \!\bigg[\frac{\pi}{2}-\epsilon\Gamma,\frac{3\pi}{2}+\epsilon\Gamma\bigg].
  \label{appeq:phase_nonlin}
\end{equation} 
The above equation reveals that the second-order $\eta$ field decreases (increases) the crest (trough) phase by $2\Gamma \epsilon$.  This will be shown to impact the estimates of $T^{\textnormal{L,crest}}$, $T^{\textnormal{L,trough}}$,  $\overline{u^{\textnormal{L,crest}}}$, and $\overline{u^{\textnormal{L,trough}}}$, however $x^{\textnormal{L,crest}}$ and $x^{\textnormal{L,trough}}$ remain unchanged.

The velocity field for the intermediate depth case reads
\begin{subequations}
\begin{align}
u(x,z,t) &= \epsilon c \dfrac{\cosh k(z+H)}{\sinh (kH)}  \cos (k x-\omega t) + \epsilon^2 \Omega c \dfrac{\cosh 2k(z+H)}{\sinh (2kH)}  \cos 2(k x-\omega t), \\
w(x,z,t)  & = \epsilon c \dfrac{\sinh k(z+H)}{\sinh (kH)}  \sin (k x-\omega t)+\epsilon^2 \Omega c \dfrac{\sinh 2k(z+H)}{\sinh (2kH)}  \sin 2(k x-\omega t),\label{eq:w_stokes2}
\end{align}
\end{subequations}
where $\Omega={3\cosh \alpha}/{(2\sinh^3\alpha)}$. The $x$--pathline equation written in terms of the Lagrangian phase is given by
\begin{equation}
\frac{d \theta}{d t}  =\omega\bigg[1-\underbrace{\epsilon \frac{\cosh (\epsilon \cos \theta+\epsilon^2 \Gamma \cos 2\theta+\alpha)}{\sinh \alpha}  \cos \theta -\epsilon^2 \Omega \frac{\cosh 2(\epsilon \cos \theta+\epsilon^2 \Gamma \cos 2\theta+\alpha)}{\sinh 2\alpha}  \cos 2\theta}_{\mathcal{X}} \bigg]. 
\label{appeq:dtheta_by_dt}
\end{equation}
We note here that the $z$-pathline equation can also be used to derive the expression for $d\theta/dt$, which is asymptotic to the expression in \eqref{appeq:dtheta_by_dt}; see Appendix A.

 Equation\,\eqref{appeq:dtheta_by_dt} on integration yields
\begin{flalign} 
  \int \! dt &  \approx \frac{1}{\omega}\!\int \!(1+\mathcal{X}+\mathcal{X}^2 +\ldots)\,d\theta 
\Rightarrow  t \!=\! \frac{\theta}{\omega}\!+\!\frac{\mathcal{C}_{\alpha} \sin\theta}{\omega}  \epsilon\!+\!\frac{(1 +\mathcal{C}_{\alpha}^2) (2\theta+\mathcal{B}\sin 2 \theta )}{4\omega}  \epsilon^2,
 \label{appeq:water_finite_2}
\end{flalign}
where $\mathcal{B}=1+\Omega/{C}_{\alpha}$ (recall that $\mathcal{B}=1$ in \eqref{eq:phase_vs_time_water}). 
Using \eqref{appeq:phase_nonlin} in \eqref{appeq:water_finite_2}, we find
the time taken by a particle under the crest and trough segments:
\begin{subequations}
\begin{align} 
 & \textrm{Crest phase duration:\,\,\,\,\,\,\,\,\,\,} T^{\textnormal{L,crest}}= \bigg[\frac{1}{2}+
   \frac{\epsilon}{\pi}(1-
\Gamma)\mathcal{C}_{\alpha}+\frac{\epsilon^2}{4}(1+\mathcal{C}_{\alpha}^2)\bigg]
   T^{\textnormal{E}},\label{eq:T_crest_general}
   \\  
 & \textrm{Trough phase duration:\,\,}T^{\textnormal{L,trough}}=
  \bigg[\frac{1}{2}-
   \frac{\epsilon}{\pi}(1-
\Gamma)\mathcal{C}_{\alpha}+\frac{\epsilon^2}{4}(1+\mathcal{C}_{\alpha}^2)\bigg]
   T^{\textnormal{E}}.
 \label{eq:T_trough_general}
\end{align}
\end{subequations}
A comparison between the analytical estimates for linear and Stokes waves and those obtained by numerically solving the respective pathline equations  in Matlab using the inbuilt ODE45 solver is given in Table \ref{tab:comparison}.

Equations \eqref{eq:T_crest_general}--\eqref{eq:T_trough_general}
reveal that $\Gamma\!>\!1$ (occurs when $\alpha \!<\! 1.196$) leads to $T^{\textnormal{L,crest}}\!<\!T^{\textnormal{L,trough}}$. Since Stokes wave theory is particularly suited for deep and intermediate water waves,  $\alpha$ shouldn't be a small number. Therefore, for a relatively small range of  intermediate $\alpha$ values,  we have $T^{\textnormal{L,crest}}\!<\!T^{\textnormal{L,trough}}$, while for $\alpha\! >\! 1.196$, the inference drawn from the linear wave theory (i.e.\,$T^{\textnormal{L,crest}}\!>\!T^{\textnormal{L,trough}}$) should hold. Additionally, we find $x^{\textnormal{L,crest}}$ and $x^{\textnormal{L,trough}}$ expressions to be identical to \eqref{eq:water_x_crest}--\eqref{eq:water_x_trough}.

The { average horizontal velocity of a particle in the crest and trough phase} are as follows:
\begin{subequations}
\begin{align}
& \textrm{{Avg. particle velocity in crest phase}:\,\,\,\,\,\,\,\,\,\,} \overline{u^{\textnormal{L,crest}}}  = \frac{2\mathcal{C}_{\alpha}}{\pi}c\epsilon+\frac{c}{2}\bigg[1+\mathcal{C}_{\alpha}^2\bigg\{1-\frac{8}{\pi^2}(1-\Gamma)\bigg\}\bigg]\epsilon^2,\label{eq:u_crest_general}\\
& \textrm{{Avg. particle velocity in trough phase}:\,\,} \overline{u^{\textnormal{L,trough}}}  =-\frac{2\mathcal{C}_{\alpha}}{\pi}c\epsilon+\frac{c}{2}\bigg[1+\mathcal{C}_{\alpha}^2\bigg\{1-\frac{8}{\pi^2}(1-\Gamma)\bigg\}\bigg]\epsilon^2. \label{eq:u_trough_general}
\end{align}
\end{subequations}
Therefore, $\forall \Gamma\in \mathbb{R}^{+}$ (for water waves, $\Gamma\geq 1/2$) we have
\begin{equation}
   |\,\overline{u^{\textnormal{L,crest}}}\,|-|\,\overline{u^{\textnormal{L,trough}}}\,|=c\bigg[1+\mathcal{C}_{\alpha}^2 \bigg\{1-\frac{8}{\pi^2}(1-\Gamma)\bigg\}\bigg]\epsilon^2>0,
    \label{eq:avg_vel_waterwave}
\end{equation}
which is consistent with the results obtained for linear waves.
 
 Summarizing this subsection (\S\ref{sec:water}\ref{Appsubsec:water}),  the velocity field due to linear waves given in \eqref{eq:pathline_x}--\eqref{eq:pathline_z} (and hence \eqref{eq:pathline_theta}) would yield second-order accurate estimates of  $T^{\textnormal{L,crest}}$, $T^{\textnormal{L,trough}}$,  $\overline{u^{\textnormal{L,crest}}}$,  $\overline{u^{\textnormal{L,trough}}}$, \emph{provided} the correct integration limits of crest and trough phases obtained from   \eqref{appeq:phase_nonlin} are used. Finally, we point out that $\Omega$ is absent in all the estimates. 

\begin{table}
  \begin{center}
\def~{\hphantom{0}}
  \begin{tabular}{lcccc}
     {}        &  \hspace{2cm} Linear wave  &  {}     & \hspace{1.5cm} Stokes wave  &  {}  \\ 
     {} & $\dfrac{T^{\textnormal{L},\textnormal{crest}}}{T^{\textnormal{E}}}$ &  \hspace{-1cm} $\dfrac{T^{\textnormal{L},\textnormal{trough}}}{T^{\textnormal{E}}}$ & \hspace{-1cm} $\dfrac{T^{\textnormal{L},\textnormal{crest}}}{T^{\textnormal{E}}}$ & \hspace{-1cm} $\dfrac{T^{\textnormal{L},\textnormal{trough}}}{T^{\textnormal{E}}}$ \\ \\
     {Theoretical value} & 0.5178 & \hspace{-1cm} 0.4848 & \hspace{-1cm} 0.5083 & \hspace{-1cm} 0.4943  \\
     {Numerical value} & 0.5179 & \hspace{-1cm} 0.4847 & \hspace{-1cm} 0.5087 & \hspace{-1cm} 0.4939  \\
  \end{tabular}
  \caption{Comparison between theoretical and numerical values of $T^{\textnormal{L},\textnormal{crest}}$ and $T^{\textnormal{L},\textnormal{trough}}$ for linear water waves and Stokes waves. Parameters: $\epsilon=0.05$, $\alpha=2$ and $T^{\textnormal{E}}=0.64$ s. 
  } 
  \label{tab:comparison}
  \end{center}
\end{table}

\subsection{Calculation of Stokes drift}
The Stokes drift velocity is given by 
\begin{equation}
\overline{u^{\textnormal{SD}}}=\frac{x^{\textnormal{L,net}}}{T^{\textnormal{L}}}=\frac{c(1+\mathcal{C}_{\alpha}^2) } {2}\epsilon^2.  
\label{eq:uSD}
\end{equation}
The above expression exactly matches with that obtained using \eqref{SD_classical}.  For completeness, we compare \eqref{eq:uSD} with the drift velocity obtained using  \eqref{eq:SD_general}:
\begin{align}
  \overline{u^{\textnormal{L}}}=\frac{1}{T^{\textnormal{L}}}\int_{0}^{T^{\textnormal{L}}} u(x(t),z(t),t) \,dt&=\frac{\epsilon}{T^{\textnormal{L}}k}\int_{0}^{2 \pi} \frac{\cosh (\epsilon \cos \theta+{\alpha})  \cos \theta }{\sinh {\alpha}-\epsilon \cosh (\epsilon \cos \theta+{\alpha})  \cos \theta} \, d\theta \nonumber \\  &=\frac{c(1+\mathcal{C}_{\alpha}^2) } {2}\epsilon^2.
  \label{eq:UL_water}
\end{align}
Similar to \eqref{eq:lag_mean_sound}, averaging in \eqref{eq:UL_water} is also performed over ${T^{\textnormal{L}}}$, and not ${T^{\textnormal{E}}}$.

The Eulerian mean $u$--velocity is given by
\begin{align}
  \overline{u^{\textnormal{E}}}=\frac{1}{T^{\textnormal{E}}}\int_{0}^{T^{\textnormal{E}}} u(x,z,t) \,dt= \frac{ c}{T^{\textnormal{E}}} \epsilon \int_{0}^{T^{\textnormal{E}}}  \frac{ \cosh (kz+{\alpha})}{\sinh {\alpha}} \cos (kx-\omega t) \,dt=0.
\end{align}
Hence  the expression obtained using $\overline{u^{\textnormal{SD}}}=\overline{u^{\textnormal{L}}}-\overline{u^{\textnormal{E}}}$  exactly matches \eqref{eq:uSD}. 

Finally we  evaluate $\overline{w^{\textnormal{SD}}}=\overline{w^{\textnormal{L}}}-\overline{w^{\textnormal{E}}}$.
\begin{align}
\overline{w^{\textnormal{L}}}&=\frac{1}{T^{\textnormal{L}}}\int_{0}^{T^{\textnormal{L}}} w(x(t),z(t),t) \,dt
=-\frac{\epsilon}{T^{\textnormal{L}}k}\int_{0}^{2 \pi} \frac{\sinh (\epsilon \cos \theta+{\alpha})  \sin \theta }{\sinh {\alpha}-\epsilon \cosh (\epsilon \cos \theta+{\alpha})  \cos \theta} \, d\theta =0.
  \label{eq:wL_water}
\end{align} 
To understand why the above integral vanishes, we need to use a substitution $\Theta=\pi-\theta$, thereby changing the integration limits from [$0,\,2\pi$] to [$-\pi,\,\pi$].  
The integrand in \eqref{eq:wL_water} is an odd function  (unlike \eqref{eq:UL_water}, where the integrand is an even function), hence must yield zero on integration in [$-\pi,\,\pi$]. Also note that the integrand is a smooth function since the  denominator does not approach zero. This is because the two terms in the denominator `$\sinh {\alpha}-\epsilon \cosh (\epsilon \cos \theta+{\alpha})  \cos \theta$' (which stems from \eqref{eq:pathline_theta}) are order separated.

 The Eulerian mean $w$--velocity is given by
\begin{align}
  \overline{w^{\textnormal{E}}}=\frac{1}{T^{\textnormal{E}}}\int_{0}^{T^{\textnormal{E}}} w(x,z,t) \,dt= \frac{ c}{T^{\textnormal{E}}} \epsilon \int_{0}^{T^{\textnormal{E}}}  \frac{ \sinh (kz+{\alpha})}{\sinh {\alpha}} \sin (kx-\omega t) \,dt=0. 
  \label{eq:wE_water}
\end{align}
This leads to the expected result $\overline{w^{\textnormal{SD}}}=0$.

It is to be noted that all expressions in \S\ref{sec:water} are for intermediate water depth. These expressions  can be straightforwardly written in the deep  ($\alpha\!\gg\!1$) and shallow  ($\alpha\!\ll\!1$) water limits by respectively substituting $\mathcal{C}_{\alpha}\!=\!1$ and $\mathcal{C}_{\alpha}\!=\!1/\alpha$.

\section{Summary and Conclusion} 

Stokes drift, i.e.\,mass transport by (specifically, water) waves in their direction of propagation, plays a central role in the transport of various oceanographic tracers.
We revisit this 175-year-old problem to answer a simple question -- what leads to Stokes drift? We undertake a  Lagrangian approach with the intention of solving the pathline equation of the generic form \eqref{eq:the_key}. In this regard, we express the particle position $\mathbf{x}$ and elapsed time $t$ in terms of the Lagrangian phase $\theta$, i.e.\, as  $\mathbf{x}(\theta)$ and $t(\theta)$. Such a parametric description provides a  unique advantage --  to probe at the `sub-wave' level, i.e., we can separately consider the crest phase $\theta \in [0,\pi/2]\cup [3\pi/2,2\pi]$ and the trough phase $\theta \in [\pi/2,3\pi/2]$, and evaluate various quantities in each of them.
Based on this, we build on the qualitative arguments of  \cite{buhler2014waves} and  \cite{van2017stokes}, and for the first time, we have mathematically shown that Stokes drift (net drift in the direction of wave motion) arises because a particle in a linear wave field spends more time,  undergoes greater horizontal displacement, and travels at a faster average horizontal velocity in the crest phase in comparison to the trough phase. We substantiate all our arguments with second-order accurate quantitative estimates obtained via asymptotic analysis. {While in \eqref{eq:exact_sound} we obtain a closed-form pathline solution $\mathbf{x}(t)$ for the specific case of linear longitudinal waves (which also naturally reveals the Lagrangian time period and Stokes drift), the approach developed in this paper that combines Lagrangian phase parameterization and asymptotic methods is more generic and meaningful.}
We note in passing that our analysis is entirely based on kinematics, and hence is different from the classical technique of obtaining particle trajectories by solving the mass and momentum conservation equations using Lagrangian coordinates \citep{lamb1932,clamond2007lagrangian}.

The classical expression given in \eqref{SD_classical} indicates that Stokes drift can occur in all kinds of non-transverse waves in fluids, which implies that the mechanism must have a generic explanation. This  motivated us to treat  the 
1D longitudinal (e.g.\,sound or shallow-water waves) and the 2D intermediate depth water waves problem under the same umbrella. We show that for both cases,  particle trajectories can be represented as a 1D autonomous dynamical system when expressed in terms of $\theta$; see \eqref{eq:sound2} and \eqref{eq:pathline_theta}. This is expected to generally hold for all kinds of waves (1D, 2D, or 3D) in fluids. Such dimensionality reduction has substantial mathematical benefits. The resulting first-order nonlinear ODE is amenable to standard asymptotic methods, using which we obtain $T^{\textnormal{L,crest}}$, $T^{\textnormal{L,trough}}$, $x^{\textnormal{L,crest}}$, $x^{\textnormal{L,trough}}$, $\overline{u^{\textnormal{L,crest}}}$, $\overline{u^{\textnormal{L,trough}}}$, $\overline{u^{\textnormal{L}}}$, 
 $\overline{u^{\textnormal{SD}}}$, etc. 
We evaluate these quantities up to $\mathrm{O}(\epsilon^2)$. Moreover, we show that finite amplitude wave theory can impact the estimates of $T^{\textnormal{L,crest}}$, $T^{\textnormal{L,trough}}$, $\overline{u^{\textnormal{L,crest}}}$, and $\overline{u^{\textnormal{L,trough}}}$, but it does not change the overall understanding obtained from the linear wave theory. We find that mathematically accurate estimates can still be obtained from the linear wave theory provided the domain of crest and trough phases take the second-order elevation effects into account. 

{While our study is restricted to irrotational flows with no underlying current, the presence of current can influence  drift velocity's magnitude and direction \citep{constantin2010pressure,gupta2021modified}.} {Moreover, the combined effect of current and sea-floor topography can also non-trivially alter the canonical Stokes drift \citep{gupta2021modified,gupta_guha_2022}.}
Wave-induced Lagrangian transport has obtained a renewed interest, specifically for a better understanding of marine pollutant transport pathways.
Recent studies \citep{santamaria2013stokes, dibenedetto2018transport,calvert2021mechanism} reveal that a particle's finite size, varied shape, and specific gravity can have important consequences. 
It is possible that a particle's shape-anisotropy and/or inertia can lead to significant differences in the transport behaviour during the crest and trough segments of a phase cycle. In such scenarios, the estimates we furnished could be beneficial. 

\clearpage
\acknowledgments
 We thank Prof.\,Leo Maas (Utrecht),  Prof.\,William  Young (Scripps), and the two anonymous reviewers for their valuable comments. 

%
%
\datastatement
Data sharing is not applicable.

%






%



\appendix[A] 
\label{App:BB}

\appendixtitle{Proof that  $d\theta/dt$ equation (\eqref{appeq:dtheta_by_dt}) derived  respectively from the $x$ and $z$ pathline equations are consistent}

We prove this consistency in a different (simpler) way. 
First, we re-write \eqref{appeq:eta} below in terms of $\theta$:
\begin{equation}
    \eta(\theta) = \frac{\epsilon}{k} \cos \theta + \frac{\epsilon^2}{k} \Gamma \cos 2 \theta.
    \label{appeq:eta_1}
\end{equation}
Next, we express $d\theta/dt$ in \eqref{appeq:dtheta_by_dt} up to $O(\epsilon)$:
\begin{equation}
\frac{d \theta}{d t}  =\omega\bigg[1- \epsilon \coth \alpha \cos \theta\bigg] + O(\epsilon^2). \label{appeq:dtheta_by_dt11}
\end{equation}
Hence at $z=\eta$, we can derive a second-order accurate expression for $dz/dt$ \emph{without} invoking the R.H.S. of \eqref{eq:w_stokes2}: 
\begin{equation}
    \frac{dz}{dt}=\frac{dz}{d\theta}\frac{d\theta}{dt}=-\epsilon c\bigg[\sin \theta +2\epsilon \Gamma \sin 2\theta\bigg]\bigg[1- \epsilon \coth \alpha \cos \theta\bigg]=-\epsilon c \sin \theta +\epsilon^2 c\bigg[\frac{\coth \alpha}{2} -2\Gamma\bigg] \sin 2\theta.
    \label{eq:key1}
\end{equation}
The goal is to show that \eqref{eq:key1} is asymptotic to  \eqref{eq:w_stokes2}. In this regard, we re-write \eqref{eq:w_stokes2} in terms of $\theta$: 
\begin{equation}
    \frac{dz}{dt}  = -\epsilon c \dfrac{\sinh (\alpha+\epsilon \cos \theta+\epsilon^2 \Gamma \cos 2\theta)}{\sinh \alpha}  \sin \theta-\epsilon^2 \Omega c \dfrac{\sinh 2(\alpha+\epsilon \cos \theta+\epsilon^2 \Gamma \cos 2\theta)}{\sinh 2 \alpha}  \sin 2\theta, \label{appeq:w_111}
\end{equation}
which yields (up to second-order accuracy)
\begin{equation}
\frac{dz}{dt}=-\epsilon c \sin \theta +\epsilon^2 c\bigg[-\frac{\coth \alpha}{2} -\Omega\bigg] \sin 2\theta.
    \label{eq:key2}
\end{equation}
Recall that $\Gamma\!=\!{(3-\tanh^2\alpha)}/{(4\tanh^3\alpha)}$ and $\Omega={3\cosh \alpha}/{(2\sinh^3\alpha)}$, which implies $2\Gamma-\Omega=\coth \alpha$. Hence proved that \eqref{eq:key1} and \eqref{eq:key2} are identical.

\appendix[B] 
\appendixtitle{$\theta$-pathline equation for an interface at an arbitrary mean water depth}
\label{app:pathline_arbit}

A material interface $\eta_i$ inside the domain at any water depth $z\!=\!-h$ (where $h\leq H$) obeys the kinematic boundary condition 
\begin{equation}
\frac{\partial \eta_i}{\partial t} = \frac{\partial \phi}{\partial z}\bigg|_{z=-h},  \label{appeq:KBC}
\end{equation}
where the velocity potential $\phi$ is evaluated from \eqref{eq:phi_wat}.
Assuming  $\eta_i=-h+a_h \cos (kx-\omega t)$, we relate the amplitude `$a_h$' of this material interface to the amplitude `$a$' of the water wave at the free surface:
\begin{equation}
    a_h=a \frac{\sinh \beta}{\sinh {\alpha}}, \label{appeq:a_as}
\end{equation}
where $\beta =\alpha-kh$.
Following the exact same procedure outlined in \S\ref{sec:water}, we derive the `$\theta$' pathline equation of a particle on the  material surface $\eta_i=-h+a_h \cos \theta$: 
\begin{equation}
    \frac{d \theta}{d t} =\omega\bigg[1-\epsilon \frac{\cosh (k a_h \cos \theta+\beta)}{\sinh \alpha}  \cos \theta\bigg],
\end{equation}
which on integration yields
\begin{align}
t= \dfrac{\theta}{\omega}+\dfrac{\epsilon}{\omega}  \frac{ \cosh \beta}{ \sinh \alpha} \sin \theta+\frac{\epsilon^2}{4 \omega}    \frac{\cosh 2\beta}{\sinh^2 \alpha}  \left({2\theta}+{\mathcal{B}\sin 2 \theta}\right),\label{eq:time_diff_water}
\end{align}
where $\mathcal{B}=1$. Similarly, following the procedure in \S\ref{sec:water}, we  obtain 
\begin{align} 
x= \dfrac{\epsilon}{k}  \frac{ \cosh \beta}{ \sinh \alpha} \sin \theta+\frac{\epsilon^2}{4k} \frac{\cosh 2\beta}{\sinh^2 \alpha} \left({2\theta}+{\mathcal{B}\sin 2 \theta}\right). \label{eq:time_diff_water}
\end{align}
Finally, the Stokes drift velocity is given as follows:
\begin{equation}
\overline{u^{\textnormal{SD}}}=\frac{x^{\textnormal{L,net}}}{T^{\textnormal{L}}}=\frac{c}{2} \frac{\cosh 2\beta}{\sinh^2 \alpha} \epsilon^2.  
\end{equation}

 \bibliographystyle{ametsocV6}
 \bibliography{references}

\begin{thebibliography}{20}
\providecommand{\natexlab}[1]{#1}
\providecommand{\url}[1]{\texttt{#1}}
\renewcommand{\UrlFont}{\rmfamily}
\providecommand{\urlprefix}{URL }
\expandafter\ifx\csname urlstyle\endcsname\relax
  \providecommand{\doi}[1]{https://doi.org/\discretionary{}{}{}#1}\else
  \providecommand{\doi}{https://doi.org/\discretionary{}{}{}\begingroup
  \urlstyle{rm}\Url}\fi
\providecommand{\eprint}[2][]{\url{#2}}

\bibitem[{Andrews and McIntyre(1978)Andrews, and McIntyre}]{andrews1978exact}
Andrews, D.~G., and M.~E. McIntyre, 1978: An exact theory of nonlinear waves on
  a {L}agrangian-mean flow. \textit{J. Fluid Mech.}, \textbf{89~(4)}, 609--646.

\bibitem[{B{\"u}hler(2014)}]{buhler2014waves}
B{\"u}hler, O., 2014: \textit{Waves and mean flows}. Cambridge University
  Press.

\bibitem[{Calvert et~al.(2021)Calvert, McAllister, Whittaker, Raby, Borthwick,,
  and Van Den~Bremer}]{calvert2021mechanism}
Calvert, R., M.~L. McAllister, C.~Whittaker, A.~Raby, A.~G.~L. Borthwick, and
  T.~S. Van Den~Bremer, 2021: A mechanism for the increased wave-induced drift
  of floating marine litter. \textit{J. Fluid Mech.}, \textbf{915}, A73.

\bibitem[{Clamond(2007)}]{clamond2007lagrangian}
Clamond, D., 2007: On the {L}agrangian description of steady surface gravity
  waves. \textit{J. Fluid Mech.}, \textbf{589}, 433--454.

\bibitem[{Constantin(2006)}]{constantin2006trajectories}
Constantin, A., 2006: The trajectories of particles in {S}tokes waves.
  \textit{Invent. Math.}, \textbf{166~(3)}, 523--535.

\bibitem[{Constantin and Strauss(2010)Constantin, and
  Strauss}]{constantin2010pressure}
Constantin, A., and W.~Strauss, 2010: Pressure beneath a {S}tokes wave.
  \textit{Comm. Pure Appl. Math.}, \textbf{63~(4)}, 533--557.

\bibitem[{Constantin and Villari(2008)Constantin, and
  Villari}]{constantin2008particle}
Constantin, A., and G.~Villari, 2008: Particle trajectories in linear water
  waves. \textit{J. Math. Fluid Mech.}, \textbf{10~(1)}, 1--18.

\bibitem[{Craik(1988)}]{craik1988wave}
Craik, A. D.~D., 1988: \textit{Wave interactions and fluid flows}. Cambridge
  University Press.

\bibitem[{DiBenedetto et~al.(2018)DiBenedetto, Ouellette,, and
  Koseff}]{dibenedetto2018transport}
DiBenedetto, M.~H., N.~T. Ouellette, and J.~R. Koseff, 2018: Transport of
  anisotropic particles under waves. \textit{J. Fluid Mech.}, \textbf{837},
  320--340.

\bibitem[{Dingemans(1997)}]{dingemans1997water}
Dingemans, M.~W., 1997: \textit{Water wave propagation over uneven bottoms},
  Vol.~13. World Scientific.

\bibitem[{Earnshaw(1860)}]{earnshaw1860}
Earnshaw, S., 1860: Viii. {O}n the mathematical theory of sound. \textit{Phil.
  Trans. R. Soc. Lond.}, \textbf{~(150)}, 133--148.

\bibitem[{Falkovich(2011)}]{falkovich2011fluid}
Falkovich, G., 2011: \textit{Fluid mechanics: A short course for physicists}.
  Cambridge University Press.

\bibitem[{Gupta and Guha(2021)Gupta, and Guha}]{gupta2021modified}
Gupta, A., and A.~Guha, 2021: Modified {S}tokes drift due to resonant
  interactions between surface waves and corrugated sea floor with and without
  a mean current. \textit{Phys. Rev. Fluids}, \textbf{6~(2)}, 024\,801.

\bibitem[{Gupta and Guha(2022)Gupta, and Guha}]{gupta_guha_2022}
Gupta, A., and A.~Guha, 2022: A new {L}agrangian drift mechanism due to
  current–bathymetry interactions: applications in coastal cross-shelf
  transport. \textit{J. Fluid Mech.}, \textbf{952}, A15.

\bibitem[{Lamb(1932)}]{lamb1932}
Lamb, H., 1932: \textit{Hydrodynamics}. University Press.

\bibitem[{Longuet-Higgins(1986)}]{longuet1986eulerian}
Longuet-Higgins, M.~S., 1986: Eulerian and {L}agrangian aspects of surface
  waves. \textit{J. Fluid Mech.}, \textbf{173}, 683--707.

\bibitem[{Santamaria et~al.(2013)Santamaria, Boffetta, Afonso, Mazzino,
  Onorato,, and Pugliese}]{santamaria2013stokes}
Santamaria, F., G.~Boffetta, M.~M. Afonso, A.~Mazzino, M.~Onorato, and
  D.~Pugliese, 2013: Stokes drift for inertial particles transported by water
  waves. \textit{Europhys. Lett.}, \textbf{102~(1)}, 14\,003.

\bibitem[{Stokes(1847)}]{stokes1847}
Stokes, G.~G., 1847: On the theory of oscillatory waves. \textit{Trans. Camb.
  Philos. Soc.}, \textbf{8}, 441.

\bibitem[{Umeyama(2012)}]{umeyama2012eulerian}
Umeyama, M., 2012: Eulerian--{L}agrangian analysis for particle velocities and
  trajectories in a pure wave motion using particle image velocimetry.
  \textit{Philos. Trans. R. Soc. A}, \textbf{370~(1964)}, 1687--1702.

\bibitem[{Van~den Bremer and Breivik(2017)Van~den Bremer, and
  Breivik}]{van2017stokes}
Van~den Bremer, T.~S., and {\O}.~Breivik, 2017: Stokes drift. \textit{Philos.
  Trans. R. Soc. A}, \textbf{376~(2111)}, 20170\,104.

\end{thebibliography}

\end{document}